\documentclass[twocolumn,showpacs,amsmath,amssymb,prd,nofootinbib]{revtex4}
\usepackage{epsfig}
\def\sss{\scriptscriptstyle}
\def\^#1{^{\sss #1}}
\def\_#1{_{\sss #1}}
\def\beq{\begin{equation}}
\def\eeqno#1{\label{#1}\end{equation}}

\def\TP{\textbf{\textsf{P}}}
\def\Pss{\textsf{P}}

\def\az{a\_{0}}

\def\l0{\ell\_{0}}

\def\rar{\rightarrow}
\def\s{\sigma}
\def\l{\lambda}

\def\f{\phi}

\def\z{\zeta}

\def\r{\rho}

\def\m{\mu}
\def\n{\nu}
\def\e{\eta}

\def\A{\mathcal{A}}

\def\L{\mathcal{L}}
\def\U{\mathcal{U}}

\def\P{\mathcal{P}}
\def\Q{\mathcal{Q}}
\def\V{\mathcal{V}}

\def\d{\delta}
\def\drt{d^3r}

\def\a{\alpha}
\def\b{\beta}
\def\c{\gamma}
\def\d{\delta}
\def\eps{\epsilon}
\def\vr{{\bf r}}
\def\vx{{\bf x}}

\def\vF{{\bf F}}

\def\vv{{\bf v}}

\def\vd{{\bf d}}

\def\va{{\bf a}}
\def\vA{{\bf A}}

\def\vF{{\bf F}}

\def\S{\Sigma}

\def\grad{\vec\nabla}
\def\div{\vec \nabla\cdot}
\def\gf{\grad\phi}

\def\RM{r\_M}

\def\gmn{g\_{\m\n}}

\def\hmn{h\_{\m\n}}

\def\gh{g^{1/2}}

\def\emn{\e\_{\m\n}}

\def\azg{\A_0}

\begin{document}
\title{General virial theorem for modified-gravity MOND}
\author{Mordehai Milgrom }
\affiliation{Department of Particle Physics and Astrophysics, Weizmann Institute}

\begin{abstract}
An important and useful relation is known to hold in two specific MOND theories. It pertains to low-acceleration, isolated systems of pointlike masses, $m_p$, at positions $\vr_p$, subject to gravitational forces $\vF_p$. It reads $\sum_p \vr_p\cdot\vF_p=-(2/3)(G\az)^{1/2}[(\sum_p m_p)^{3/2}-\sum_p m_p^{3/2}]$; $\az$ is the MOND acceleration constant. Here I show that this relation holds in the nonrelativistic limit of {\it any} modified-gravity MOND theory. It follows from only the basic tenets of MOND, which include departure from standard dynamics at accelerations below $\az$, and space-time scale invariance in the nonrelativistic, low-acceleration limit. This implies space-dilatation invariance of the static, gravitational-field equations, which, in turn, leads to the above point-mass virial relation. Thus, the various MOND predictions and tests based on this relation hold in any modified-gravity MOND theory. Since we do not know that any of the existing MOND theories point in the right direction, it is important to identify such predictions that hold in a much larger class of theories. Among these predictions are the MOND two-body force for arbitrary masses, and a general mass-velocity-dispersion relation of the form  $\s^2=(2/3)(MG\az)^{1/2}[1-\sum_p (m_p/M)^{3/2}]$, where $M=\sum_p m_p$.
\end{abstract}
\pacs{}
\maketitle
\section{\label{introduction} Introduction}
MOND is a paradigm of dynamics that departs significantly  from Newtonian dynamics (ND) and general relativity (GR) at low accelerations, such as those characterizing galactic systems. MOND was put forth \cite{milgrom83} to account for the mass discrepancies in the Universe without dark matter (DM) and, possibly, without ``dark energy''. Reference \cite{fm12} is an extensive review of MOND.
\par
The basic premises of MOND are the following: (1) dynamics in galactic systems, and the universe at large, involve a new fundamental constant with the dimensions of acceleration, $\az$, (2) at high accelerations, much above $\az$ -- i.e., when we take the formal limit $\az\rar 0$ -- standard dynamics is restored, and (3) in the limit of low acceleration $\ll\az$ -- the deep-MOND limit (DML) -- nonrelativistic (NR) dynamics become space-time scale invariant (SI).
\par
In addition, for a theory to qualify as a MOND theory, it must describe test-particle dynamics that asymptotically far from a bounded mass, $M$, depend only on $M$, and not on how this mass is distributed \cite{milgrom14} (see Sec. \ref{virial} for more details).
\par
Some major predictions of MOND -- in the form of general laws -- follow from these basic tenets alone; e.g., the asymptotic flatness of rotation curves and the mass-asymptotic-speed relation (see a detailed account in Ref. \cite{milgrom14}). Some of these laws pertain to phenomena involving the transition from Newtonian to DML dynamics at accelerations $\sim\az$. Others concern disparate phenomena in DML systems.  These latter phenomena, which concern us here, are predicted to be governed by SI dynamics. It follows from SI that it is always possible to write the DML theory such that $\az$ and $G$ do not appear separately, only in the product $\azg\equiv G\az$ (see below, and, for more details, e.g., Refs. \cite{milgrom09a,milgrom14}).
\par
In many instances of testing MOND, one uses it to predict virial-equilibrium velocities of galactic systems from the observed (baryonic) masses in the system, and compare these with the measured velocities. Many of these systems are ``pressure-supported'' (or ``random'') systems, such as dwarf-spheroidal and elliptical galaxies, binary galaxies, galaxy groups,  etc.
\par
A central tool for applying the above procedure to such systems is a MOND relation between the system mass, $M$, and some measure of its mean velocity dispersion, $\s$. It follows from the basic tenets alone \cite{milgrom14} that for DML systems the ratio $Q\equiv\s^2/(M\azg)^{1/2}$ is independent of the mass and size of a system; it can depend only on dimensionless attributes of the system, such as mass ratios of subcomponents, shape parameters, anisotropy ratios, etc. Furthermore, $Q$ has to be of order of magnitude of unity. This may serve as a rough tool for predicting $\s$ from $M$.
\par
However, for meaningful testing and predictions we need a more concrete and accurate determination of $Q$ and its dependence on system parameters. An exact result was obtained \cite{milgrom97,milgrom10a} using the specific forms of two MOND theories: the nonlinear Poisson version \cite{bm84}, and quasi-linear MOND (QUMOND) \cite{milgrom10a}. In both, it was shown that for an {\it isolated} DML system of pointlike masses, $m_p$, at positions $\vr_p$, subject to gravitational forces $\vF_p$, we have
 \beq \V_{pm}\equiv -\sum_p \vr_p\cdot\vF_p=\frac{2}{3}\azg^{1/2}[M^{3/2}-\sum_p m_p^{3/2}], \eeqno{i}
 where $M=\sum_p m_p$; the quantity
$\V_{pm}$ may be termed the ``point-mass virial''.
This is a powerful result with various applications.  In particular, it gives the exact expression for $\Q$ for any isolated, DML system of pointlike masses in a steady-state equilibrium, for which relation (\ref{i}) leads to \cite{milgrom94,milgrom97},
\beq Q=\frac{2}{3}[1-\sum_p (m_p/M)^{3/2}],  \eeqno{jarat}
provided $\s^2=M^{-1}\sum_p m_p\vv_p^2$ is the mass-weighted mean squared velocity in the system.\footnote{Velocities within the bodies do not enter, only the bodies' center-of-mass velocities.} To wit, $Q$ depends only on the mass ratios of the constituent bodies, not on any other dimensionless attribute (many of which are much harder, if not impossible, to determine observationally).
When the constituents can be considered test particles, $\sum_p (m_p/M)^{3/2}\ll 1$, and $Q \approx 2/3$ is universal.
\par
Such relations have been used to test MOND, e.g., in small galaxy groups where the galaxies are not test masses \cite{milgrom98,milgrom02a}, and in the dwarf-spheroidal satellites of the Andromeda Galaxy \cite{mm13,mm13a}. Another result of relation (\ref{i}) is the DML, two-body force between arbitrary masses, used, for example, in the recent study of the Milky Way-Andromeda system \cite{zhao13}. Yet another application is in the definition of a reduced $Q$ parameter  as applied to disc galaxies, which may be useful in discriminating between ``modified-gravity'' (MG) and ``modified-inertia'' formulations of MOND \cite{milgrom12b}.
Additional applications are discussed in Refs. \cite{milgrom97,milgrom10a}.
The frequent use of relation (\ref{i}) had been based on its emergence from only the two specific formulations of MOND we have today.
\par
I show here, most significantly, that relation (\ref{i}) is a prediction of the NR limit of all (relativistic) MG MOND theories. It follows as a DML result from only the basic tenets applied to the general form of such theories (with the help of a few additional reasonable assumptions).
This means that its past and future predictions and applications hold in a much larger class of theories than thought before.
This is particularly welcome, since we do not know that any of the presently known MOND theories point in the right direction; so it is helpful to identify MOND predictions that are less theory dependent.
\par
In Sec. \ref{mg}, I define MG MOND theories and discuss some of their properties.
In Sec. \ref{virial}, I derive relation (\ref{i}) for the NR, DML limit of this general class of theories.
Section \ref{discussion} is a summary and discussion, where, in particular, I give a step-by-step outline of the derivation, which also summarizes the assumptions that enter. This summary can be consulted as a road map for the derivation, which is somewhat lengthy.
\section{\label{mg}Modified-gravity MOND theories}
A relativistic, MG MOND theory is a metric theory where the matter action, $S\_M$, is the standard one, with matter coupling to the metric in the standard, minimal way. The Einstein-Hilbert action of GR is replaced by a modified action, $S\_G(\gmn,A,c,G,\az)$, which may involve additional gravitational degrees of freedom (DOFs) -- not coupled directly to matter -- of arbitrary tensorial character, marked collectively $A$. $S\_G$ may involve higher derivatives, or be nonlocal; it may be a general functional of the gravitational DOFs. This class of theories includes the relativistic formulations of MOND known today: TeVeS \cite{bekenstein04}, MOND adaptations of Einsten-Aether theories \cite{zlosnik07}, bimetric MOND (BIMOND) theories \cite{milgrom09}, and nonlocal metric theories \cite{deffayet11}.
I consider purely gravitational systems; so $S\_M$ is the standard particle action $S\_M=-c^2\sum_pm_p\int d\tau_p$.
\par
In the NR limit of this theory (the approximation for near Minkowskian space time: $\gmn=\emn+\hmn$, $|\hmn|\ll 1$, and slow motions),\footnote{Any acceptable theory should have a well-defined ``slow-motion'', or ``static'', limit, where the solutions exist, and are unique (given the appropriate boundary conditions), when the mass distribution is assumed static, and all time derivatives are taken to vanish. I take the MOND theory under study to have such  a limit.} the action becomes $S=\int L~dt$, where, in the continuum description of the matter mass distribution, $\r(\vr)$,
 \beq L=\int\r(\vr)[\frac{1}{2}\vv^2(\vr)-\f(\vr)]\drt-
L_f(\f,\psi,G,\az). \eeqno{kalur}
The first two terms come from the matter action, and are common to all MG theories considered here. In principle, $L_f$ may be a general functional of the gravitational DOFs, but for the sake of concreteness I specialize to Lagrangians of the form
 \beq  L_f=\int  \L_f(\f,\psi,G,\az)~\drt.  \eeqno{bufa}
Here, as usual, $\f=-c^2h\_{00}/2$ is the NR gravitational potential, and $\psi\_a$ are the other NR, gravitational DOFs -- called  collectively $\psi$ -- such as are coming from other elements of the metric, and those descending from $A$. They can be of any (Euclidean) tensorial rank, and $\L_f$ is a functional of $\f$ and $\psi\_a$ that does not involve the time in our NR approximation (e.g., no time derivatives); it involves $G$ and $\az$ as the only dimensioned constants.
\par
Under an infinitesimal change in a DOF, say of $\psi\_a$, we have for the change in $L_f$ coming from a volume $v$
 \beq \d L_f=\int_v\frac{\d L_f}{\d\psi\_a} \d\psi\_a~\drt+\int\_{\S}\vec\U_a(\f,\psi,\d\psi\_a)\cdot\vec{d\s}, \eeqno{gret}
where $\S$ is the surface of $v$; $\U_a$ is a functional homogeneous of degree 1 in $\d\psi\_a$; and similarly for a variation on $\f$. Then, it is posited that solutions of the theory are those that annihilate $\d L_f$ for any $\d\psi\_a$ and $\d\f$ that annihilate the surface integral.
The equations of motion (EOMs) are then
 \beq \dot\vv=-\gf, ~~~~~\frac{\d L_f}{\d\f}=-\r.~~~~~ \frac{\d L_f}{\d\psi\_a}=0.  \eeqno{manud}
For general $L_f$, not necessarily of the form (\ref{bufa}), we need to apply this procedure to the whole space; so eq.(\ref{gret}) is written with $v$ the whole space, and $\S$ the surface at infinity.
\par
We shall hereafter concentrate on the ``potential energy'' $L\_V=\int\r\f~\drt+L_f$, the extremization of which over $\f,~\psi$ describes the static problem whereby the gravitational fields are determined from $\r$, treated as a given external source.  $\r$ is, of course, determined from the masses making up the system, and their positions, which are additional DOFs of the general problem (\ref{manud}). But, as usual, the field problem can be solved separately from the motion of the masses in the NR approximation.
\par
Consider the DML of the NR theory; so let us assume that $L_f$ already takes its DML form. To constitute a MOND theory, the EOMs (\ref{manud}) must then be space-time SI. In particular, SI of the first equation tells us that under
$(t,\vr)\rar\l(t,\vr)$ we have $\f(\vr)\rar\f(\vr/\l)$; so the scaling dimension of $\f$ is zero. The scaling dimension of the other DOFs is defined such that if $\psi\_a=V\^{i\_1...i\_K}\_{j\_1...j\_N}$ is a tensor, and transforms under scaling as $\psi\_a(\vr)\rar\l\^{K-N+\a\_a}\psi\_a(\vr/\l)$, then $\a\_a$ is the scaling dimension of $\psi\_a$.
\par
Without loss of generality, we can assume that all the DOFs have dimensions that match their scaling dimensions; i.e., if $[\psi\_a]=[m]^\b[l]^\c[t]^\zeta$ then $\a\_a=\c+\zeta$. Otherwise, we can normalize $\psi\_a$ by a power of $\az$ that will lead to this ($\f$ is already standardized). With this standardized choice, SI implies that $G$ and $\az$ cannot appear in the problem except as $\azg=\az G$ (see, e.g., Ref. \cite{milgrom14}); so $\L_f=\L_f(\f,\psi,\azg)$.
\par
Since $L_f$ describes a static system, SI implies that the 2nd and 3rd eq. (\ref{manud}) are invariant to space dilatations $\vr\rar\l\vr$, if the scaling dimensions of $\f$ and $\psi\_a$ are taken to be also their dilatation dimensions (I use ``scaling'' for space-time, and ``dilatation'' for space only).
\par
The first term in $L\_V$ is clearly invariant to space dilatations, under which $\r(\vr)\rar\l^{-3}\r(\vr/\l)$, since $\f$ has zero dimension. And, $L_f$, like $\r\f$, must have dimension $-3$ under dilatations. Namely, when all DOFs are transformed as described above, $\L_f(\vr)\rar \l^{-3}\L_f(\vr/\l)$. Thus, $\L_f(\vr)~\drt\rar \L_f(\vr/\l)d^3(\vr/\l)$, and so, by change of integration variable we see that for infinitesimal dilatations $\l=1+\eps$, $L_f$ changes by a surface integral over the surface $\S$ of $v$:
\beq\d L_f=(\int_{v/\l}-\int_{v})\L_f(\vr)~\drt\approx
-\eps\int\_\S\L_f \vr\cdot\vec{d\s}. \eeqno{mulsa}
This implies, as required, that the EOMs are dilatation invariant. However, $L_f$ itself is not quite invariant: asymptotically far from the masses, the system becomes spherically symmetric, and $\L_f$ depends on $r$ only; so it has to behave as $\L_f\propto r^{-3}$ to have the correct dilatation transformation. $L_f$ itself thus diverges logarithmically, and despite its formal invariance under dilatations, is subject to a finite change, as the surface term is finite.

\subsection{Forces on bodies}
A body is the collection of masses within a subvolume $v$ that does not overlap with other masses. The gravitational force acting on the body
is
 \beq \vF_v=-\int_v\r\gf~\drt, \eeqno{forza}
since its center-of-mass acceleration is $\vA_v=M_v^{-1}\vF_v$, with the mass of the body $M_v=\int_v\r~\drt$. It can be seen (e.g., Ref. \cite{milgrom02}) that the force generates the change in $L\_V$ due to infinitesimal, rigid translations of the body by $\vec\eps$: $\d L\_V=-\vec\eps\cdot\vF_v$.
\par
It is useful to consider the stress tensor, $\TP$, analogous to the energy-momentum tensor of the gravitational action $L_f$, associated with the gravitational DOFs, defined standardly as follows: write a covariant version, $L^c_f$, of $L_f$, on a curved-space background with a metric $g\_{ij}$, with derivatives becoming covariant derivatives $\drt\rar \gh\drt$, $\d\_{ij}\rar g\_{ij}$, etc. Then, $\TP$ is defined such that under an infinitesimal change $g\_{ij}\rar g\_{ij} +\d g\_{ij}$ ($\d g\_{ij}$ vanishes fast enough at infinity),
 \beq \d L^c_f\equiv\frac{1}{2}\int \gh \Pss^{ij}\d g\_{ij}~\drt. \eeqno{lipo}
Then, in $\TP$ we take back $g\_{ij}\rar \d\_{ij}$ to get the Euclidean value (understood hereafter) of $\TP$.
\par
When all the DOFs that appear in the action used to define $\TP$ (other than the metric) do not appear elsewhere in the total action,  $\TP$ is conserved (i.e., divergenceless) for solutions of the EOMs (``on shell'').\footnote{This follows from the fact that the action is a coordinate scalar (and using the EOMs).} But this is not the case here, since $\f$ appears also in the $\int\r\f$ part of the action.\footnote{We can include this part (whose covariant form does not involve the metric anyway) in the definition of $\TP$, but then the appearance of $\r$ in it, which contains matter DOFs, should be reckoned with, leading to the same result.}
So, following the standard arguments for showing that the energy-momentum tensor is divergenceless, here we find, instead, for the divergence of $\TP$,
 \beq \div\TP=-\r\gf~~~~~(\Pss^{ij}_{,j}=-\r\d\^{ik}\f\_{,k}).   \eeqno{nuter}
This result is not related to MOND and follows only from the way $\f$ appears in the Lagrangian of the form (\ref{kalur}).
\par
Using relation (\ref{nuter}) in expression (\ref{forza}), the force on a body can be written as a surface integral
 \beq \vF_v=\int_v \div\TP~\drt=\int\_\S\TP\cdot \vec{d\s}, \eeqno{aua}
where $\S$ is any closed surface containing the body and no other mass.
\par
Importantly, because $L_f$ is dilatation invariant (in the DML), it can be shown that for solutions of the EOMs
 \beq \P\equiv {\rm Tr}(\TP)=\d\_{ij}\Pss^{ij}=\partial_i \U^i(\f,\psi), \eeqno{katura}
Namely, the trace of the stress tensor is a divergence of some vector functional of the fields, when these solve the EOMs.
Not committing ourselves to Lagrangians of the form (\ref{bufa}), we have the weaker result that $\int \P~\drt$ can be written as a surface integral at infinity of a functional of $\f$ and the $\psi$s. This will suffice for our purpose. For conformally invariant theories, we further have $\P=0$.
\par
Equation (\ref{katura})
is a well known result for scale- and conformally invariant field theories (e.g., Ref. \cite{nakayama13}), but it is worth explaining how it emerges in the present context: as we saw, under dilatations
$\f(\vr)\rar\f(\vr/\l)$,  $\psi_a=V\^{i\_1...i\_K}\_{j\_1...j\_N}(\vr)\rar \l\^{K-N+\a\_a}\psi_a(\vr/\l)$ (so every $\vr$ derivative is multiplied by $1/\l$), we have $\L_f(\vr)\rar \l^{-3}\L_f(\vr/\l)$. This implies, in turn, that $\l$ disappears altogether if in the curved-space form of $L_f$, we replace $\f(\vr)\rar\f(\vr)$,  $\psi_a=V\^{i\_1...i\_K}\_{j\_1...j\_N}(\vr)\rar \l\^{\a\_a}\psi_a(\vr)$,\footnote{Note that the factor $\l\^{K-N}$ is not included here, and that the independent variable $\vr$ is not scaled.} $g\_{ij}(\vr)\rar\l^{2}g\_{ij}(\vr)$, $g\^{ij}(\vr)\rar\l^{-2}g\^{ij}(\vr)$,
$\gh\rar\l^3\gh$, namely: $\gh\L_f(g\_{ij},\f,\psi\_a)(\vr)\rar\l^3\gh\L_f(\l^2g\_{ij},\f,\l^{\a\_a}\psi_a)(\vr)=
\gh\L_f(g\_{ij},\f,\psi\_a)(\vr)$. This is because all tensorial and covariant-derivative indices are contracted either among themselves or with the metric.
Consider then an infinitesimal transformation of the latter form with $\l=1+\eps$ under which $L_f$ does not vary (since the independent variable $\vr$ is not changed now). By the definition of $\TP$, eq.(\ref{lipo}), and eq.(\ref{gret}) (and remembering that $\d\f=0$) we have
$$0=\d L_f=\frac{1}{2}\int_v \gh \Pss^{ij}\d g\_{ij}~\drt+$$
 \beq \int_v\sum_a\frac{\d L_f}{\d\psi\_a} \d\psi\_a~\drt+ \int\_{\S}\sum_a\vec\U_a(\f,\psi,\d\psi\_a)\cdot\vec{d\s}. \eeqno{nerq}
 In our case $\d\psi\_a=\eps\a\_a\psi\_a$, $\d g\_{ij}=2\eps g\_{ij}$. So using the EOM, we get, after taking the Euclidean limit,
 \beq \int_v \P~\drt=\int\_\S\vec\U\cdot\vec{d\s},   \eeqno{cupi}
 where
  \beq \vec\U=-\sum_a\a\_{a}\vec\U_a(\f,\psi,\psi\_a). \eeqno{pidra}
Inasmuch as this holds for any volume, we have
  \beq \P=\div\vec\U.  \eeqno{mama}
More generally, eq.(\ref{cupi}) holds for the whole space and $\S$ is a surface at infinity.
\par
In a conformally invariant theory there is invariance to the above transformation with $\l(\vr)$ an arbitrary function of $\vr$. So now $\d g\_{ij}=2\eps(\vr) g\_{ij}$. Applying eq.(\ref{nerq}) to the whole space, and taking $\eps(\vr)$ that vanishes fast enough at infinity but is arbitrary elsewhere, we have $\int \P(\vr)\eps(\vr)~\drt=0$; so, $\P(\vr)=0$, for solutions of the EOMs.
\par
In the nonlinear Poisson formulation of MOND we have
$\P=0$, and, indeed, the static-gravity part of the theory is conformally invariant \cite{milgrom97}. In QUMOND $\P=\div\vec\U\not=0$, but $\vec\U$ decreases faster than $r^{-2}$ at infinity, so $\int\_{space} \P~\drt=0$ \cite{milgrom10a}.

\section{\label{virial}The virial relation}
We start with the ``continuum virial''\footnote{So termed to distinguish it from the ``point-mass'' virial, of which it is the continuum limit, or the limit where all the masses may be considered as test particles.}
\beq \V\equiv \int\r\vr\cdot\gf~\drt \eeqno{rapula}
(integration is over the whole space), defined for an isolated, self-gravitating system.
$\V$ does not depend on the choice of origin, since shifting the origin by $\vr_0$ changes $\V$, by $\vr_0\cdot\vF$, where $\vF$ is the total force on the system and vanishes.
$\V$ was calculated explicitly, for systems in the DML, in the nonlinear Poisson formulation \cite{milgrom97} and in QUMOND \cite{milgrom10a}, where it was found that  in both
\beq\V=\frac{2}{3}M^{3/2}\azg^{1/2}.   \eeqno{virta}
Our main step in this paper is the realization that this is true, in fact, for the general class of theories we consider here, and that it follows only from the basic MOND tenets.
\par
To see this, use the general relation (\ref{nuter}) to write, integrating by parts,
\beq \V=-\int r\^i\d\_{ik}\Pss^{kj}_{,j}~\drt=\int\P~\drt-
\int\_{\S\_{\infty}}\vr\cdot\TP\cdot\vec{d\s}.\eeqno{tadre}
Using eq.(\ref{cupi}), which rests on the dilatation invariance,
 \beq \V=\int\_{\S\_{\infty}}(\vec\U-\vr\cdot\TP)\cdot\vec{d\s}.   \eeqno{yuta}
Thus, in DML theories, $\V$ can be written as an integral over the surface at infinity over some functional of $\f$ and $\psi\_a$.\footnote{This can also be shown by considering directly the variation of $L\_V$ under a dilatation of the source $\r(\vr)\rar\l^{-3}\r(\vr/\l)$ ($\l=1+\eps$), under which all fields are dilatation transformed. On one hand, this change is given by the surface term in eq.(\ref{mulsa}), on the other hand it is given by the virial plus surface terms from eq.(\ref{gret}).} This is not true in ND.
\par
As stated in Sec. \ref{introduction}, it is required of an MG MOND theory that $\f$ depend asymptotically only on $M$, and thus do not have a preferred direction.
\par
Since $\f$ is of dilatation dimension zero, it must behave asymptotically as ${\rm ln}(r)$. Dimensional considerations dictate that it must be $\f\propto(M\azg)^{1/2}{\rm ln}(r)$. The normalization of $\az$ is defined such that
 \beq \f=(M\azg)^{1/2}{\rm ln}(r). \eeqno{majara}
\par
It is possible, in principle -- especially in higher-derivative theories, in which vacuum solutions are characterized by more integration constants -- for the large-radius behavior of $\f$ to depend on details of the mass distribution. But, in an acceptable MOND theory we require this dependence to decay faster than the leading logarithm. Otherwise, for example, we do not get a sharp mass-asymptotic-speed relation, and we exclude such theories from the outset. SI of the DML does, in itself, imply plausibly that size characteristics of the system must be asymptotically subdominant,\footnote{Seen by noting that if $R$ sets the scale of the size, then asymptotically $\f$ can depend only on $R/r$.} but shape information may, in principle, enter the asymptotic behavior, which our added assumption disallows.
\par
I stretch this condition somewhat and posit that the dominant asymptotic contribution to the surface integral in eq.(\ref{yuta}) also depends on $M$ only, and not on details of its distribution.\footnote{Once $\f$ is known, $\psi_a$ are determined from only the third set of eq.(\ref{manud}); so clearly if $\f$ has to be spherically symmetric so are all other DOFs. While this makes our extended assumption plausible, it does not prove it, since the nonsphericity of $\f$ in the inner regions may, in principle, produce asymptotic nonsphericity in the $\psi_a$ and the surface integrand.} Then, dimensional arguments
dictate that $\V=k M^{3/2}\azg^{1/2}$. The coefficient $k$ is fixed as follows: consider a system made of a mass $M$ bounded in a volume $v$, and a test (negligible) mass $m$ at position $\vr\_m$, in a very small volume, far from $v$. The virial for the whole system, $\V=k(M+m)^{3/2}\azg^{1/2}$ can also be calculated by taking $\gf$ to be solely due to $M$, since $m$ is a test particle. The integral in eq.(\ref{rapula}) thus has a contribution from $v$, which is just the virial for $M$ alone, i.e. $kM^{3/2}\azg^{1/2}$, and that from integrating over the small volume of $m$, $m\vr\_m\cdot\gf(\vr\_m)$.
But $\f$ must already have its asymptotic form (\ref{majara}) at $\vr\_m$; so this latter contribution is $mM^{1/2}\azg^{1/2}$. Thus $k(M+m)^{3/2}= kM^{3/2}+mM^{1/2}+o(m/M)$, giving $k=2/3$, yielding the generality of eq.(\ref{virta}).
\par
Relation (\ref{i}), which I want to derive, follows from eq.(\ref{virta}) in the way described briefly in Ref. \cite{milgrom97}. Here, I recap the argument, stating more clearly the underlying assumptions and approximation.
Consider a system $\r(\vr)$ that can be separated into nonoverlapping bodies of masses $m_p$. The extent of body $p$ is $e_p$, defined, say, as the radius of the smallest sphere containing the whole body and centered at its center of mass $\vr_p$.  Another relevant radius is $\RM^p=(m_pG/\az)^{1/2}$, the MOND radius of $m_p$, far beyond which (from $m_p$) its influence is in the DML.
\par
Relation (\ref{i}) applies to DML systems of pointlike masses, $m_p$, defined by the requirements that (a) both $e_p$ and $\RM^p$ are much smaller than all the separations $r_{pq}$ of the body from the rest, and (b) everywhere far outside all the spheres of radii $\RM^p$ we have $|\gf|\ll\az$.
\par
It follows from eq.(\ref{aua}) that $\vF_p$ can be written as a surface integral over any closed surface, $\S$, enclosing $m_p$ alone. From the above conditions follows that we can choose $\S$ to be of order $r_{pq}$ in extent, i.e., much larger than both $e_p$ and $\RM^p$. Our basic assumption then tells that this integral, and thus $\vF_p$, does not depend on the way $m_p$ is distributed within $e_p$, as long as its $e_p\ll r_{pq}$. Starting then with a system satisfying the above definition of a DML, pointlike system, we can replace it, {\it without changing the point-mass virial}, with a standardized system in which (a) $e_p\gg\RM^p$, so the body itself is in the DML, but, (b) the field within the body still dominates over the field due to the rest of the system at its position. For example, if this latter field is $\eta\az$ ($\eta\ll 1$), and $\RM^p=\z e_p$, this requires that $\eta\ll\z\ll 1$.\footnote{We can then smear the mass smoothly within this $e_p$ so that the system is everywhere in the DML.}
\par
To recapitulate, we have erected a standardized system (a) that has the same masses, positions, and point-mass virial as the original one, (b) that is in the DML everywhere, so we can use our DML results above, and (c) where within each mass, its own field strongly dominates over the correction to the field due to the other masses.\footnote{Masses in the original system may violate conditions b or c. For example, a mass can have $e_p\ll\RM^p$; so is not itself in the DML, or it can be so large for its mass that the external field it is in dominates its own. These, however, are not obstacles.} We now calculate the point-mass virial for this standardized system.
\par
The continuum virial, $\V$, can be written as a sum of integrals over the bodies:
\beq \V=\sum_p \int_{v_p}\r\vr\cdot\gf~\drt. \eeqno{ragasa}
Each term is not the virial produced by body $p$ alone, since $\gf$ is produced by the whole system.
For each body write $\vr=\vr_p+\vx$ ($|\vx|\le e_p\ll r_{pq}$), so the $p$th term is  $-\vr_p\cdot\vF_p+\int_{v_p}\r\vx\cdot\gf~d^3x$. Now, let $\gf_p$ be the field that would have been produced by body $p$ if it were alone, and write $\gf=\gf_p+\gf_p\^{ex}$ ( $\gf_p\^{ex}$ is not the field produced alone by the other bodies, since we are dealing with a nonlinear theory). Thus $\int_{v_p}\r\vx\cdot\gf~d^3x=
\int_{v_p}\r\vx\cdot\gf_p~d^3x+
\int_{v_p}\r\vx\cdot\gf_p\^{ex}~d^3x$.
The second term vanishes in the point-mass limit $e_p/r_{pq}\rar 0$.
The first term, however, does not vanish in the limit, since in the DML
$\gf_p\sim 1/|\vx|$. This integral is, in fact, the continuum virial of $m_p$ when alone, and equals
$\V_p=(2/3)m_p^{3/2}\azg^{1/2}$.  We thus end up with the required expression (\ref{i}) for the ``point-mass virial.''
\par
The choice of pointlike masses in a given system is not unique. If our system is a group of galaxies, for example, we may choose the galaxies, or we may choose the stars in these galaxies, as the pointlike constituents. The point-mass virial relation (\ref{i}) holds for either, provided our assumptions are satisfied for them (so in the second case the galaxies have to be DML stellar systems in themselves, not only the group as a system of galaxies). In applications, such as eq.(\ref{jarat}), the choice of constituents enter both the right-hand side, through the list of masses, and the left-hand side through the definition of $\s$ that enters $Q$. If galaxies in a group are our masses, then velocities of the stars within them do not enter $\s$ (only the galaxies' center-of-mass velocities do).
\par
More generally, note that we have not made any assumption on the internal dynamics of the constituents, which may even be governed by forces other then gravity (such as if they are atoms or molecules). Only the masses' contributions to the general gravitational field enter.

\section{\label{discussion}Summary and discussion}
I showed that the very useful point-mass virial relation (\ref{i}) is a prediction of any MG theory that satisfies the basic tenets of MOND, plus some plausible assumptions, not related to MOND in particular. The arguments and assumptions leading to this result are as follows:
\begin{enumerate}
\item
One starts by restricting the discussion to MG theories. This means that the dynamics of matter is governed by a metric. One further restricts to purely gravitational systems; so masses are the only constants characterizing matter. In the NR limit, which one further restricts to, this pinpoints the single gravitational potential, $\f$, as determining the dynamics of masses (via $\va=-\gf$), hence the special role of the virial, which is defined using $\f$ alone. The special role of $\f$ in MG theories also singles it out in the expression for the divergence of the stress tensor, eq.(\ref{nuter}).
\item
The virials (continuum and point-mass), $\V,~\V_{pm}$, are scale invariant quantities since $\gf$ and $\vF=m\va$ scale as $\l^{-1}$. So, they can be written as functions of scale-invariant attributes of the system. This is true in MOND as well as in ND.
\item
In MOND, only $\az$ is allowed as additional constant, and thus, in the DML, assumed SI by the basic tenets, only $\azg$ appears.
 \item
$\V$ and  $\V_{pm}$ have the same dimensions as $M^{3/2}\azg^{1/2}$, which is invariant under scaling of the space time units; so the ratio $\V M^{-3/2}\azg^{-1/2}$ is dimensionless and scale invariant. It can thus depend only on mass ratios, length ratios (shape parameters), etc.
In ND, the same is true of the ratio $\V R/MG$, where $R$ is some size characteristic of the system. The striking fact about the MOND case is, however, that unlike the ND case (where $\V=\langle\vr\cdot\gf\rangle$), the virials do not depend on any shape parameter only on the constituent masses. Furthermore, the exact dependence on the masses can be derived, and is simple.
\item
In the DML, SI implies dilatation invariance of the static gravitational-field equations -- not shared by ND. This leads to the continuum virial being writable as a surface integral at infinity (which, indeed, does not hold in ND).
\item
Then enters the assumption that the theory is such that the asymptotic fields, and hence the expression for the continuum virial, are dominated by the contribution that depends only on the total mass. It follows that for the continuum virial, $\V M^{-3/2}\azg^{-1/2}$ is a constant of the theory, independent on any dimensionless attributes of the system.
\item
The normalization of the virial is then fixed by normalizing $\az$ to give eq.(\ref{majara}).
\item
In the final step we generalized to the case of a system of poinlike bodies of finite masses, assuming that these are much smaller than their separations. We then showed that the continuum virial of the whole system is the sum of the required point-mass virial and the individual continuum virials of all the masses, considered each as being alone.
\end{enumerate}
\par
Relation (\ref{i}) is exact in the simultaneous limits evident from our derivation: the DML limit, and the pointlike limit, in which extents of bodies are much smaller than separations. Otherwise, the relation is the lowest-order result in these small parameters.
In real systems we expect corrections of order constituents size over separations, which presumably depend on various dimensionless system parameters such as shape parameters and details of the mass distribution (e.g., mass ratios).
\par
Another important result of relation (\ref{i}), beside eq.(\ref{jarat}), is the general DML (attractive) two-body force for two masses $m_1,~m_2$, a distance $\ell$ apart \cite{milgrom94,milgrom97}:\footnote{Yet other applications are, e.g., an expression for the (inward) force per unit length of a ring of mass $M$ and radius $R$:
$F=M^{3/2}\azg^{1/2}/3\pi R^2$. The force per unit area of a thin spherical shell of mass $M$ and radius $R$ is $F=M^{3/2}\azg^{1/2}/6\pi R^3$.}
 \beq F(m_1,m_2,\ell)=\frac{2}{3}\frac{\azg^{1/2}}{\ell}[(m_1+m_2)^{3/2}-m_1^{3/2}-m_2^{3/2}]. \eeqno{twobody}
\par
It is instructive to check all the above in the more general class of theories \cite{milgrom10a,milgrom14}, whose DML $\L_f$ is of the form
 \beq \L_f=\azg^{-1}\sum_{a,b}s\_{ab} [(\gf)^2]^\eta[(\grad\psi)^2]^\xi(\gf\cdot\grad\psi)^\theta, \eeqno{iiaa}
where $\eta=a+3/2$, $\xi=a+b(2-p)/2$, $\theta= b(p-1)-2a$;
 $p$ is fixed for a given theory, and $a,~b$ are arbitrary.
The dimensions of $\f$ and $\psi$ are, respectively, $[l]^{2}[t]^{-2}$ and, if $b\not= 0$, $[l]^{2-p}[t]^{2(p-1)}$ (for $b=0$, the dimensions of $\psi$ are arbitrary); $s\_{ab}$ are dimensionless. For any $p$, this reduces to the
nonlinear Poisson theory for $a=b=0$. QUMOND is obtained for $p=-1$ with two terms with $a=-3/2,~b=1$  and $a=-b=-3/2$. For $p=0$ we have $\P=0$ for any combination of $a,~b$, and the DML is conformally invariant. Likewise for $b=0$, in which case $p$ does not enter.\footnote{Many of these theories may be unfit for various reasons.}
\par
There may be MOND theories that allow gravitating masses of opposite signs, as, e.g., in BIMOND with twin matter \cite{milgrom10b}. So we can have systems with vanishing total mass. In the asymptotic regime of these, $\f$ is not radial, is not logarithmic, and does depend on details of the mass distribution (much like higher multipole fields, which dominate asymptotically for a system of vanishing total charge in Maxwellian electrostatics). Dilatation of $\r(\vr)$ itself does affect the asymptotic field. For example, in NR BIMOND, which may be the nonlinear Poisson equation, or QUMOND -- depending on the version of BIMOND at hand -- the field equation was solved exactly in Ref. \cite{milgrom10b} for a DML system of two opposite pointlike masses $\pm m$. Asymptotically, the potential is $\f\approx -(m\azg)^{1/2}\vr\cdot\vd/r^2$,
where $\vd$ is the dipole separation. So, asymptotic speeds decrease as $(m\azg)^{1/4}(d/r)^{1/2}$. The breakdown of the general result occurs because the asymptotic potential is not invariant to the dilatation of the mass distribution (under which $\vd\rar \l\vd$).
Our result for the point-mass virial still holds in this case, with $\sum_p m_p=0$, since the fields decay fast enough for the surface integral in eq.(\ref{yuta}) to vanish, so the continuum virial for the whole system vanishes, still satisfying eq.(\ref{virta}).
\par
In any event, since masses of opposite signs repel each other, we cannot have a self-gravitating system of this type, so we do not expect to find such a galactic system.


\begin{thebibliography}{}
\bibitem{milgrom83}M. Milgrom, Astrophys. J. 270, 365 (1983).
\bibitem{fm12}B. Famaey and S. McGaugh, Living Rev. Relativity 15, 10 (2012).
\bibitem{milgrom14}M. Milgrom, Mon. Not. R. Astron. Soc. 437, 2531 (2014).
\bibitem{milgrom09a}M. Milgrom, Astrophys. J. 698, 1630 (2009).
\bibitem{milgrom97}M. Milgrom, Phys. Rev. E 56, 1148 (1997).
\bibitem{milgrom10a}M. Milgrom, Mon. Not. R. Astron. Soc. 403, 886 (2010).
\bibitem{bm84}J. Bekenstein and M. Milgrom, Astrophys. J. 286, 7 (1984).
\bibitem{milgrom94}M. Milgrom, Astrophys. J. 429, 540 (1994).
\bibitem{milgrom98}M. Milgrom, Astrophys. J. Lett. 496, L89 (1998).
\bibitem{milgrom02a}M. Milgrom, Astrophys. J. Lett. 577, L75 (2002).
\bibitem{mm13}S. McGaugh and M. Milgrom,  Astrophys. J. 766, 22 (2013).
\bibitem{mm13a}S. McGaugh and M. Milgrom, Astrophys. J. 775, 139 (2013).

\bibitem{zhao13}H.S. Zhao {\it et al.}, Astron. Astrophys. Lett. 557, L3 (2013).
\bibitem{milgrom12b}M. Milgrom, Phys. Rev. Lett. 109, 251103 (2012).
\bibitem{bekenstein04}J.D. Bekenstein, Phys. Rev. D 70  083509  (2004).
\bibitem{zlosnik07}T.G. Zlosnik, P.G. Ferreira, and G.D. Starkman, Phys. Rev. D 75 044017 (2007).
\bibitem{milgrom09}M. Milgrom, Phys. Rev. D 80, 123536 (2009).
\bibitem{deffayet11}C. Deffayet, G. Esposito-Farese, and R.P. Woodard, Phys. Rev. D 84, 124054 (2011).
\bibitem{milgrom02}M. Milgrom, J. Phys. A 35, 1437 (2002).
\bibitem{nakayama13}Y. Nakayama, arXiv:1302.0884 (2013).
\bibitem{milgrom10b}M. Milgrom, Mon. Not. R. Astron. Soc. 405, 1129 (2010).
\end{thebibliography}
\end{document}